\documentclass[12pt]{article}

\begin{document}

\pagestyle{empty}

\noindent {\small USC-98/HEP-B2\hfill \hfill hep-th/9804028}\newline
{\small \hfill }

{\vskip 0.7cm}

\begin{center}
{\Large Conformal Symmetry and Duality between}

{\Large Free Particle, H-atom and Harmonic Oscillator}$^{*}$ \\[0pt]

{\vskip 0.5cm}

\textbf{Itzhak Bars} {\Large \ \\[0pt]
}

{\vskip 0.4cm}

\textbf{Department of Physics and Astronomy}

\textbf{University of Southern California}

\textbf{Los Angeles, CA 90089-0484}

{\vskip 0.5cm}

\textbf{ABSTRACT}
\end{center}

\noindent {\small We establish a duality between the free massless
relativistic particle in }$d${\small \ dimensions, the non-relativistic
hydrogen atom (}$1/r${\small \ potential) in }$(d-1)${\small \ space
dimensions, and the harmonic oscillator in }$(d-2)${\small \ space
dimensions with its mass given as the lightcone momentum of an additional
dimension. The duality is in the sense that the classical action of these
systems are gauge fixed forms of the same worldline gauge theory action at
the classical level, and they are all described by the same unitary
representation of the conformal group SO}$(d,2)${\small \ at the quantum
level. The worldline action has a gauge symmetry }$Sp(2)${\small \ which
treats canonical variables }$(x,p)${\small \ as doublets and exists only
with a target spacetime that has d spacelike dimensions and two timelike
dimensions. This spacetime is constrained due to the gauge symmetry, and the
various dual solutions correspond to solutions of the constraints with
different topologies. For example, for the H-atom the two timelike
dimensions }$X^{0^{\prime }},X^0$ {\small live on a circle. The model
provides an example of how realistic physics can be viewed as existing in a
larger covariant space that includes two timelike coordinates, and how the
covariance in the larger space unifies different looking physics into a
single system.}

\vfill \hrule width 6.7cm \vskip 2mm

{$^{*}$ {\small Research partially supported by the US. Department of Energy
under grant number DE-FG03-84ER40168.}}

\vfill\eject

\setcounter{page}1\pagestyle{plain}

\section{Gauge sectors and duality}

In a recent paper \cite{dualconf} a duality was constructed between several
simple physical systems by showing that they are different aspects of the
same quantum theory. The theory is based on gauging the Sp$(2)$ duality
symmetry that treats position and momentum $(x,p)$\ as a doublet in phase
space. The worldline action has a manifest SO$\left( d,2\right) $ symmetry
acting linearly on a target spacetime $X^M\left( \tau \right) $ with two
times. Thanks to the gauge symmetry the theory is equivalent to a theory
with a single time, but the choice of ``time'' is not unique. For different
gauge choices of ``time'' the Hamiltonian looks different and appears to
describe different physical systems. However these systems are gauge
equivalent, i.e. duality equivalent. It was shown that the Sp$\left(
2\right) $ duality \textit{gauge invariant sector} is fully characterized in
the quantum theory by a \textit{unique} unitary representation of the
conformal group SO$\left( d,2\right) $. The quadratic Casimir of SO$\left(
d,2\right) $ takes the value $C_2=1-d^2/4$ and all higher Casimirs $C_n$ are
also fixed. In \cite{dualconf} it was shown that the free relativistic
particle is described by this representation at the quantum level. In this
paper we will show that the Hydrogen atom and the harmonic oscillator are
also described by the same unitary representation and hence they are dual to
the free particle at the classical as well as quantum levels.

First we give the action for the model. To remove the distinction between
position and momentum we rename them $X_{1}^{M}\equiv X^{M}$ and $%
X_{2}^{M}\equiv P^{M}$ and define the doublet $X_{i}^{M}=\left(
X_{1}^{M},X_{2}^{M}\right) .$ The local Sp$\left( 2\right) $ acts as follows
\begin{equation}
\delta _{\omega }X_{i}^{M}\left( \tau \right) =\varepsilon _{ik}\omega
^{kl}\left( \tau \right) X_{l}^{M}\left( \tau \right) .  \label{doublet}
\end{equation}
Here $\omega ^{ij}\left( \tau \right) =\omega ^{ji}\left( \tau \right) $ is
a symmetric matrix containing three local parameters, and $\varepsilon _{ij}$
is the Levi-Civita symbol that is invariant under Sp$\left( 2,R\right) $ and
serves to raise or lower indices. The Sp$\left( 2,R\right) $ gauge field $%
A^{ij}\left( \tau \right) $ is symmetric in $(ij)$ and transforms in the
standard way $\delta _{\omega }A^{ij}=\partial _{\tau }\omega ^{ij}+\omega
^{ik}\varepsilon _{kl}A^{lj}+\omega ^{jk}\varepsilon _{kl}A^{il}.$ The
covariant derivative is $D_{\tau }X_{i}^{M}=\partial _{\tau
}X_{i}^{M}-\varepsilon _{ik}A^{kl}X_{l}^{M}.$ An action that is invariant
under this gauge symmetry is
\begin{eqnarray}
S_{0} &=&\frac{1}{2}\int_{0}^{T}d\tau \left( D_{\tau }X_{i}^{M}\right)
\varepsilon ^{ij}X_{j}^{N}\eta _{MN}  \label{action} \\
&=&\int_{0}^{T}d\tau \left( \partial _{\tau }X_{1}^{M}X_{2}^{N}-\frac{1}{2}%
A^{ij}X_{i}^{M}X_{j}^{N}\right) \eta _{MN}\,\,.  \nonumber
\end{eqnarray}
As argued in \cite{dualconf} this system exists non-trivially only if $\eta
_{MN}$ has signature $\left( d,2\right) $ including two timelike dimensions.
Thus there is a manifest global SO$\left( d,2\right) $ symmetry. The
canonical conjugates are $X_{1}^{M}=X^{M}$ and $\partial S/\partial \dot{X}%
_{1}^{M}=X_{2}^{M}=P^{M}$. They are consistent with the idea that $%
(X_{1}^{M},X_{2}^{M})$ is the doublet $\left( X^{M},P^{M}\right) $. There
has been some discussion in the past of an action related to this one \cite%
{marnelius}-\cite{followup}, but not including our non-trivial classical and
quantum solutions or our point view on duality. This action can be
generalized in several ways consistently with the Sp$\left( 2\right) $ gauge
invariance, including supersymmetry, and interactions with background
gravitational fields $G_{(MN)}$ and $B_{[MN]}$ and/or background gauge
fields $A_{i}^{M}$ that are doublets of Sp$\left( 2\right) $ \cite{dualconf}%
. In the presence of background fields the global symmetry SO$\left(
d,2\right) $ is replaced by the Killing symmetries of the background fields.

The equations of motion for $X_{i}^{M},A^{ij}$ that follows from the
Lagrangian \ref{action} are
\begin{eqnarray}
\left(
\begin{array}{c}
\partial _{\tau }X^{M} \\
\partial _{\tau }P^{M}%
\end{array}
\right) &=&\left(
\begin{array}{cc}
A^{12} & A^{22} \\
-A^{11} & -A^{12}%
\end{array}
\right) \left(
\begin{array}{c}
X^{M} \\
P^{M}%
\end{array}
\right)  \label{motion} \\
X\cdot X &=&X\cdot P=P\cdot P=0.  \label{constraints}
\end{eqnarray}
The global symmetry generators for SO$\left( d,2\right) $ are
\begin{equation}
L^{MN}=\varepsilon ^{ij}X_{i}^{M}X_{j}^{N}=X^{M}P^{N}-X^{N}P^{M}.
\label{lmn}
\end{equation}
They are manifestly Sp$\left( 2\right) $ \textit{gauge invariant}. At the
classical level all Casimirs of SO$\left( d,2\right) $ vanish due to the
constraints (\ref{constraints})
\begin{equation}
C_{n}\left( SO\left( d,2\right) \right) =\frac{1}{n!}Tr\left( iL\right)
^{n}=0,\quad classical.  \label{casclass}
\end{equation}
It was shown that this is sufficient to characterize completely all the
classical solutions without making any gauge choice for ``time'' \cite%
{dualconf}.

When the theory is quantized and orders of operators $X,P$ are taken into
account, there is a similar statement. Before taking the constraints (\ref%
{constraints}) into account, the quadratic Casimir of the gauge group is
\begin{equation}
C_{2}(Sp\left( 2\right) )=\frac{1}{4}\left[ X^{M}P^{2}X_{M}-\left( X\cdot
P\right) \left( P\cdot X\right) +\frac{d^{2}-4}{4}\right] ,
\end{equation}
where the last term results from operator reordering. For $L^{MN}$ of the
form (\ref{lmn}) all the Casimirs $C_{n}\left( SO\left( d,2\right) \right) $
of eq.(\ref{casclass}) can all be written in terms of the quadratic Casimir
of the gauge group $C_{2}(Sp\left( 2\right) )$ plus operator reordering
constants that depend on $d$. In particular
\begin{equation}
C_{2}\left( SO\left( d,2\right) \right) =\frac{1}{2}L_{MN}L^{MN}=\left[
C_{2}(Sp\left( 2\right) )+1-\frac{d^{2}}{4}\right] .  \label{c2}
\end{equation}
In the gauge invariant sector the physical states are singlets of Sp$\left(
2\right) $ and therefore for physical states
\begin{equation}
C_{2}(Sp\left( 2\right) )=0,\quad C_{2}\left( SO\left( d,2\right) \right) =1-%
\frac{d^{2}}{4}.
\end{equation}
Similarly all $C_{n}$ are fixed at the quantum level by demanding $%
C_{2}(Sp\left( 2\right) )=0$. Thus, the quantum solution of the theory
corresponds to a unique unitary representation of SO$\left( d,2\right) ,$
with specific eigenvalues of the Casimirs $C_{n}\left( d\right) $. This
important information obtained in covariant quantization completely
determines the unitary physical Hilbert space. There are no ghosts in the
physical space because this SO$\left( d,2\right) $ representation is unitary.

The physical content of the system is better understood in non-covariant
quantization by choosing a ``time'' and constructing a Hamiltonian. The
choice of time is not unique because the spacetime of our model has more
than one timelike dimension $X^{0^{\prime }},X^0$. Since there is more than
one ``time'' there are different looking Hamiltonians that are canonically
conjugate to the given choice of time. In such physical gauges the system is
automatically unitary but one must verify that quantization is consistent
with the global SO$\left( d,2\right) $ symmetry. This requires some
non-trivial ordering of canonical operators in the construction of the
\textit{gauge invariant} quantum generators $L^{MN}$ expressed in fixed
gauges. After doing so, we show that the unitary representation is identical
to the one that emerged from covariant quantization, with the same Casimir
eigenvalues $C_n\left( d\right) $, but now expressed in the physical basis
of some Hamiltonian. Examples of this procedure include the relativistic
massless particle, the H-atom, the harmonic oscillator, and more.

\section{ Free particle and H-atom as gauge choices}

Consider the basis $X^{M}=\left( X^{+^{\prime }},X^{-^{\prime
}},X^{+},X^{-},X^{i}\right) $ with the metric $\eta ^{MN}$ taking the values
$\eta ^{+^{\prime }-^{\prime }}=\eta ^{+-}=-1$ in the lightcone type
dimensions, while $\eta ^{ij}=\delta ^{ij}$ for the remaining $d-2$ space
dimensions. Thus one time $X^{0^{\prime }}$ is a linear combination of $%
X^{\pm ^{\prime }},$ and the other $X^{0}$ is a linear combination of $%
X^{\pm }$. The gauge group Sp$\left( 2\right) $ has three gauge parameters,
hence we can make three gauge choices. The free particle lightcone gauge is $%
X^{+^{\prime }}=1$, $P^{+^{\prime }}=0$, $X^{+}=\tau $. This is a legitimate
gauge choice as shown in \cite{dualconf}. Inserting this gauge into the
constraints (\ref{constraints}), and solving them, one finds the following
components expressed in terms of the remaining independent degrees of
freedom $(x^{-},p^{+},\vec{x}^{i}\,,\vec{p}^{i})$%
\begin{eqnarray}
M &=&\left[ +^{\prime }\quad ,\quad \quad -^{\prime }\quad \quad \quad
,\quad \quad +\quad ,\quad -\quad ,\quad i\,\,\,\,\,\right]  \nonumber \\
X^{M} &=&[1,\,\,\,\,\,\quad (\vec{x}^{2}/2-\tau x^{-})\,\,\,\,\,\,,\quad
\quad \,\tau \quad \,\,,\quad x^{-},\quad \vec{x}^{i}\,\,]
\label{relparticle} \\
P^{M} &=&[0,\,\,(\vec{x}\cdot \vec{p}-x^{-}p^{+}-\frac{\tau \vec{p}^{2}}{%
2p^{+}}),\,\,\,\,\,p^{+},\,\,\,\,\frac{\vec{p}^{2}}{2p^{+}}%
\,\,\,\,,\,\,\,\,\,\,\,\vec{p}^{i}].  \nonumber
\end{eqnarray}
One can verify that this gauge corresponds to the free relativistic massless
particle, by inserting the gauge fixed form (\ref{relparticle}) into the
action (\ref{action}). Since all constraints have been solved, the $A^{ij}$
terms are absent, and we get
\begin{eqnarray}
S_{0} &=&\int_{0}^{T}d\tau \,\,\partial _{\tau }X_{1}^{M}X_{2}^{N}\,\eta
_{MN}  \nonumber \\
&=&\int_{0}^{T}d\tau \,\,\left( \partial _{\tau }\vec{x}\cdot \vec{p}%
-\partial _{\tau }x^{-}p^{+}-\frac{\vec{p}^{2}}{2p^{+}}\right) .
\end{eqnarray}
This is the action of the free massless relativistic particle in the
lightcone gauge, in the first order formalism, with the correct Hamiltonian $%
P^{-}=\vec{p}^{2}/2p^{+}.$ Note that both time coordinates have been gauge
fixed, $X^{+^{\prime }}=1$ and $X^{+}=\tau $, to describe the free particle.
This is the free particle ``time''. The SO$\left( d,2\right) $ symmetry
generators in this gauge were given in \cite{dualconf}. They will be used in
section 4 to discuss the duality between the free massless particle and the
harmonic oscillator.

We now show that the Hydrogen atom corresponds to another gauge choice in
this system, with a rather different choice of \textquotedblleft
time\textquotedblright\ as a function of the two timelike dimensions $%
X^{0^{\prime }},X^{0}$. Consider the basis $X^{M}=\left( X^{0^{\prime
}},X^{0},X^{I}\right) $ and $P^{M}=\left( P^{0^{\prime }},P^{0},P^{I}\right)
$ with metric $\eta ^{0^{\prime }0^{\prime }}=\eta ^{00}=-1$ and $\eta
^{IJ}=\delta ^{IJ}$. Choose one gauge such that the four functions $%
X^{0^{\prime }},X^{0},P^{0^{\prime }},P^{0}$ are expressed in terms of three
functions $F,G,u$
\begin{eqnarray}
X^{0^{\prime }} &=&F\cos u,\quad X^{0}=F\sin u  \label{Htt0} \\
P^{0^{\prime }} &=&-G\sin u,\quad P^{0}=G\cos u.
\end{eqnarray}%
Inserting this form in the constraints (\ref{constraints}) gives
\begin{eqnarray}
X^{M} &=&F\left[ \cos u,\,\,\sin u\,\,,\,\,\,n^{I}\right] \\
P^{M} &=&G\left[ -\sin u,\,\cos u\,\,\,,\,\,\,m^{I}\right] ,
\end{eqnarray}%
where $n^{I},m^{I}$ are \textit{euclidean} unit vectors that are orthogonal.
We choose the following parametrization for these unit vectors in the basis $%
I=\left[ 1^{\prime },i\right] $ where $I=1^{\prime }$ denotes the extra
space dimension and $i=1,2,\cdots ,(d-1)$ labels ordinary space,
\begin{eqnarray}
n^{I} &=&\left[ -\frac{1}{\alpha }\sqrt{-2H}\,\mathbf{r\cdot p,\quad }(\frac{%
1}{r}\mathbf{r}^{i}-\frac{\mathbf{r\cdot p}}{\alpha }\mathbf{p}^{i}\mathbf{)}%
\right] ,  \label{nm} \\
m^{I} &=&\left[ \,\quad (1-\frac{r\mathbf{p}^{2}}{\alpha })\quad ,\quad
\sqrt{-2H}\,\frac{r}{\alpha }\mathbf{p}^{i}\mathbf{\,\,\,\,}\right] ,
\nonumber
\end{eqnarray}%
where
\begin{equation}
H=\frac{\mathbf{p}^{2}}{2}-\frac{\alpha }{r},  \label{HatomH}
\end{equation}%
is the Hydrogen atom Hamiltonian. This gives a solution of the constraints (%
\ref{constraints}) that have taken the form $n^{I}n^{I}=m^{I}m^{I}=1$ and $%
m^{I}n^{I}=0$. Along with these we use the freedom of choosing two more
gauge functions. One gauge choice is
\begin{equation}
GF=\frac{\alpha }{\sqrt{-2H}},  \label{Hgauge}
\end{equation}%
and the last gauge choice is a gauge for \textquotedblleft
time\textquotedblright\
\begin{equation}
u\left( \tau \right) =\left( \mathbf{r\cdot p}-2\tau H\right) \frac{\sqrt{-2H%
}}{\alpha }.  \label{Htime}
\end{equation}%
Note that time $\tau $ is embedded in $X^{0^{\prime }},X^{0}$ in a rather
complicated way given through (\ref{Htt0}, \ref{Hgauge}, \ref{Htime}).

To verify that this gauge choice really corresponds to the H-atom, we insert
it in the action (\ref{action}) and verify that it reduces to the action for
the H-atom. Since all the constraints are explicitly solved, the $A^{ij}$
terms drop out and we get
\begin{eqnarray}
S_{0} &=&\int_{0}^{T}d\tau \,\,\partial _{\tau }X_{1}^{M}X_{2}^{N}\,\eta
_{MN}  \nonumber \\
&=&\int_{0}^{T}d\tau \,\,GF\left( -\partial _{\tau }u+m^{I}\partial _{\tau
}n^{I}\right)  \label{Haction} \\
&=&\int_{0}^{T}d\tau \,\,\left( \mathbf{p}^{i}\partial _{\tau }\mathbf{r}^{i}%
\mathbf{-}H\right) .  \nonumber
\end{eqnarray}%
To derive the second line we only use the form of Eq.(\ref{Htt0}) and the
the fact that $m^{I},n^{I}$ are unit vectors. To derive the third line we
have used the explicit gauge choices (\ref{Hgauge},\ref{Htime}) to compute
the following expressions
\begin{eqnarray}
\left( GF\right) m^{I}\partial _{\tau }n^{I} &=&\mathbf{r\cdot p}\partial
_{\tau }\ln \sqrt{-2H}-\partial _{\tau }\left( \mathbf{r\cdot p}\right) +%
\mathbf{p}\cdot \partial _{\tau }\mathbf{r}, \\
-\left( GF\right) \partial _{\tau }u &=&-\mathbf{r\cdot p}\partial _{\tau
}\ln \sqrt{-2H}-\partial _{\tau }\left( \mathbf{r\cdot p-}3\tau H\right) -H.
\nonumber
\end{eqnarray}%
Also a total derivative $\partial _{\tau }\left( \mathbf{-}3\tau H-2\mathbf{%
r\cdot p}\right) $ has been dropped since it does not contribute to the
action.

The last form of the action (\ref{Haction}) is the first order formalism,
with the H-atom Hamiltonian given in (\ref{HatomH}). This form shows that
the unconstrained variables $(\mathbf{r}^{i},\mathbf{p}^{i}$) are the
standard canonical variables. The middle line of (\ref{Haction}) shows that
the H-atom in $\left( d-1\right) $ space dimensions has SO$\left( d\right) $
symmetry. The first line shows that the H-atom has a dynamical symmetry SO$%
\left( d,2\right) $ which mixes the two timelike coordinates with the $d$
space coordinates.

Through the two explicit examples discussed in this section, we have
illustrated that the Sp$\left( 2\right) $ gauge covariant action is capable
of describing not only the free particle but also complicated systems like
the H-atom, and others. The underlying reason for this is the ability to
choose time as a gauge in non-unique ways because we have more than one
timelike coordinate in the $d+2$ dimensional spacetime. For each choice of
time embedded in $X^{0},X^{0^{\prime }}$ the corresponding canonical
Hamiltonian looks different. Nevertheless these special systems are Sp$%
\left( 2\right) $ gauge equivalent, or dual to each other.

\section{SO$\left( d,2\right) $ and the H-atom}

The SO$\left( d,2\right) $ symmetry generators in the H-atom gauge are
obtained by inserting the gauge (\ref{Htt0}-\ref{Htime}) in the gauge
invariant $L^{MN}$. In a Hamiltonian formalism before ordering operators, we
have at general $\tau $
\begin{eqnarray}
L^{0^{\prime }I} &=&\frac{\alpha }{\sqrt{-2H}}\left( m^{I}\cos u+n^{I}\sin
u\right) ,\quad L^{0I}=\frac{\alpha }{\sqrt{-2H}}\left( m^{I}\sin
u-n^{I}\cos u\right)  \\
L^{0^{\prime }0} &=&\frac{\alpha }{\sqrt{-2H}},\quad L^{IJ}=\frac{\alpha }{%
\sqrt{-2H}}\left( n^{I}m^{J}-n^{J}m^{I}\right) .
\end{eqnarray}%
By inserting the forms of $n^{I},m^{I}$ given in the previous section one
can verify that the SO$\left( d\right) $ subgroup has generators $%
L^{IJ}=(L^{ij},L^{1^{\prime }i})$ that are interpreted as angular momentum
and the Runge-Lenz vector generalized to any dimension $d$
\begin{equation}
L^{ij}=\mathbf{r}^{i}\mathbf{p}^{j}-\mathbf{r}^{j}\mathbf{p}^{i},\quad
L^{1^{\prime }i}=\frac{\alpha }{\sqrt{-2H}}\left( \frac{1}{2}L^{ij}\mathbf{p}%
_{j}+\frac{1}{2}\mathbf{p}_{j}L^{ij}-\alpha \frac{\mathbf{r}^{i}}{r}\right) .
\end{equation}%
These SO$\left( d\right) $ generators are already written in their quantum
ordered and hermitian form (the factor $\sqrt{-2H}$ commutes with the
Runge-Lenz vector and can be written on either side). The classical version
of $L^{0^{\prime }i},L^{01^{\prime }}$ are (not quantum ordered)
\begin{eqnarray}
L^{0^{\prime }i} &=&\cos \left( \left( \mathbf{r\cdot p}-2\tau H\right)
\frac{\sqrt{-2H}}{\alpha }\right) r\mathbf{p}^{i}  \label{L0i} \\
&&+\frac{\alpha }{\sqrt{-2H}}\sin \left( \left( \mathbf{r\cdot p}-2\tau
H\right) \frac{\sqrt{-2H}}{\alpha }\right) (\frac{\mathbf{r}^{i}}{r}-\frac{%
\mathbf{r\cdot p}}{\alpha }\mathbf{p}^{i}\mathbf{)}\quad   \nonumber \\
L^{0^{\prime }1^{\prime }} &=&\frac{\alpha }{\sqrt{-2H}}\cos \left( \left(
\mathbf{r\cdot p}-2\tau H\right) \frac{\sqrt{-2H}}{\alpha }\right) (1-\frac{r%
\mathbf{p}^{2}}{\alpha }) \\
&&-\sin \left( \left( \mathbf{r\cdot p}-2\tau H\right) \frac{\sqrt{-2H}}{%
\alpha }\right) \mathbf{r\cdot p}  \nonumber \\
L^{0i} &=&\sin \left( \left( \mathbf{r\cdot p}-2\tau H\right) \frac{\sqrt{-2H%
}}{\alpha }\right) r\mathbf{p}^{i}\mathbf{\,\,} \\
&&-\frac{\alpha }{\sqrt{-2H}}\cos \left( \left( \mathbf{r\cdot p}-2\tau
H\right) \frac{\sqrt{-2H}}{\alpha }\right) (\frac{\mathbf{r}^{i}}{r}-\frac{%
\mathbf{r\cdot p}}{\alpha }\mathbf{p}^{i}\mathbf{)}  \nonumber \\
L^{01^{\prime }} &=&\frac{\alpha }{\sqrt{-2H}}\sin \left( \left( \mathbf{%
r\cdot p}-2\tau H\right) \frac{\sqrt{-2H}}{\alpha }\right) (1-\frac{r\mathbf{%
p}^{2}}{\alpha })  \label{L01} \\
&&+\cos \left( \left( \mathbf{r\cdot p}-2\tau H\right) \frac{\sqrt{-2H}}{%
\alpha }\right) \,\mathbf{r\cdot p}
\end{eqnarray}%
One can verify that all of these $L^{MN}$ are constants of motion, as
expected from the fact that they are symmetries of the action. Namely after
using the equations of motion for $\mathbf{r}\left( \tau \right) ,\mathbf{p}%
\left( \tau \right) $ given by \ref{Haction}, one finds $dL^{MN}/d\tau $ for
any $\tau $.

The quantum ordering of these operators must be consistent with the
commutators involving $L^{0^{\prime }0}$ and the quantum ordered generator $%
L^{1^{\prime }i}$ given above. Another consistency requirement on the
quantum ordering is that $\left( L^{0^{\prime }0},L^{0^{\prime }1^{\prime
}},L^{01^{\prime }}\right) $ must form an SO$\left( 1,2\right) $ algebra.
The representation of the conformal group SO$\left( d,2\right) $ appropriate
for the H-atom can then be discussed. We will not do this here, but rather
we will choose another gauge below where the ordering is much simpler but
yet non-trivial. However, in the present gauge, an important observation at
the quantum level is that the quadratic Casimirs of the SO$\left( d\right) $
and SO$\left( 2\right) $ subgroups are both related to the H-atom
Hamiltonian as follows (watching orders of operators)
\begin{eqnarray}
C_{2}\left( SO\left( d\right) \right)  &=&\left( L^{1^{\prime }i}\right)
^{2}+\frac{1}{2}\left( L^{ij}\right) ^{2} \\
&=&\frac{\alpha ^{2}}{-2H}-\frac{1}{4}\left( d-2\right) ^{2}, \\
\left( L^{0^{\prime }0}\right) ^{2} &=&\frac{\alpha ^{2}}{-2H}.
\end{eqnarray}%
Therefore, the SO$\left( d,2\right) $ basis labelled by the subgroups $%
|SO\left( d\right) ,SO\left( 2\right) >$ is of special interest since the
representation consists of all the quantum states of the H-atom taken
together in a single irreducible representation. At a fixed energy level the
SO$\left( d\right) $ subgroup explains the degeneracies. This is analogous
to the well known SO$\left( 4\right) $ symmetry in three dimensions. The
generators $L^{0^{\prime }1^{\prime }},L^{0^{\prime }i},L^{01^{\prime
}},L^{0i}$ mix different energy levels.

There is another construction for SO$\left( d,2\right) $ that is simpler for
discussing the H-atom at the quantum level. We will discuss the quantum
theory in more detail in the other basis, since we want to also show that
there is another approach to find the H-atom, without going through the
arguments for the choice of time at the classical level given above. In the
second approach we simply choose some generator of SO$\left( d,2\right) $
and call it \textquotedblleft Hamiltonian\textquotedblright . This approach
is also simpler for finding the harmonic oscillator as one of the dual
sectors. For the H-atom to emerge it is evident from the discussion above
that diagonalizing the generator $L^{00^{\prime }}$ would be of interest.
This will be done below.

To discuss the second approach to the H-atom we choose another gauge. In the
basis $M=\left( +^{\prime },-^{\prime },0,i\right) $, where $i=1,2,\cdots
,\left( d-1\right) ,$ we make the three gauge choices (at $\tau =0$)$,$ $%
X^{+^{\prime }}=0,\,P^{+^{\prime }}=1$ and $P^{0}=0$, and then solve the
three constraints. The result is\footnote{%
The relativistic massless particle of eq.(\ref{relparticle}) can also be
described in the timelike gauge in the basis $M=\left( +^{\prime },-^{\prime
},0,i\right) ,$ at $\tau =0$ we have
\begin{eqnarray}
X^{M} &=&\left( 1,\frac{\mathbf{r}^{2}}{2},0,\mathbf{r}^{i}\right) \\
P^{M} &=&\left( 0,\mathbf{r\cdot p},\left| \mathbf{p}\right| ,\mathbf{p}%
^{i}\right) .
\end{eqnarray}
The second H-atom gauge (\ref{Hsec}) is related to this particle gauge by a
discrete Sp$\left( 2\right) $ duality transformation that interchages $X^{M}$
and $P^{M}$ and then renaming $\mathbf{r\longleftrightarrow p}$.}
\begin{equation}
X^{M}=\left( 0,\mathbf{r\cdot p},r,\mathbf{r}^{i}\right) ,\quad P^{M}=\left(
1,\frac{\mathbf{p}^{2}}{2},0,\mathbf{p}^{i}\right) .  \label{Hsec}
\end{equation}
The canonical operators $(\mathbf{r,p)}$ in this gauge should not be
identified with the canonical operators in the other H-atom gauge (\ref{Htt0}
- \ref{Htime}), they are not simply related. The generators of the conformal
group SO$\left( d,2\right) $ are $L^{MN}=X^{M}P^{N}-X^{N}P^{M}$. They are
invariant under the Sp$\left( 2\right) $ gauge transformations, therefore
they can be evaluated in any gauge. Inserting our gauge fixed form we obtain
(recall $\eta ^{+^{\prime }-^{\prime }}=\eta ^{00}=-1$)
\begin{eqnarray}
L^{-^{\prime }+^{\prime }} &=&L_{+^{\prime }-^{\prime }}=\frac{1}{2}\left(
\mathbf{r\cdot p+p\cdot r}\right) \\
L^{0+^{\prime }} &=&L_{0-^{\prime }}=r\quad \quad \\
L^{0-^{\prime }} &=&L_{0+^{\prime }}=\frac{1}{2}\mathbf{p}^{i}r\mathbf{p}%
^{i}+\frac{a}{r} \\
L^{i+^{\prime }} &=&L_{-^{\prime }i}=\mathbf{r}^{i} \\
L^{i-^{\prime }} &=&L_{+^{\prime }i}=-\frac{1}{2}\mathbf{p\cdot \,r\,\,p}%
^{i}-\frac{1}{2}\mathbf{p}^{i}\mathbf{\,r\cdot p\,\,}+\frac{1}{2}\mathbf{p}%
^{j}\mathbf{r}^{i}\mathbf{p}^{j}+b\frac{\mathbf{r}^{i}}{\mathbf{r}^{2}} \\
L^{i0} &=&L_{0i}=-\frac{1}{2}\left( r\mathbf{p}^{i}+\mathbf{p}^{i}r\right) \\
L^{ij} &=&L_{ij}=\mathbf{r}^{i}\mathbf{p}^{j}-\mathbf{r}^{j}\mathbf{p}^{i}
\end{eqnarray}
The system is quantized according to the standard commutation relations
\begin{equation}
\left[ \mathbf{r}^{i},\mathbf{p}^{j}\right] =i\delta ^{ij},
\end{equation}
and all operators are ordered to insure that all components of $L^{MN}$ are
hermitian. In general there are ordering ambiguities. These are denoted by
the constants $a$ and $b$ that appear in $L_{+^{\prime }0}$ and $%
L_{+^{\prime }i}$. For example, consider the classical expression for $%
L_{+^{\prime }0}=\frac{1}{2}\mathbf{p}^{2}r$\textbf{. }There are several
possible quantum orderings, all of which are hermitian, and all of which are
consistent with rotation symmetry. For example, for any parameter $\lambda $
we have a hermitian operator ordered as $\frac{1}{2}\mathbf{p}%
^{2}r\rightarrow \frac{1}{2}r^{\lambda }\mathbf{p}^{i}r^{1-2\lambda }\mathbf{%
p}^{i}r^{\lambda }$. This may be reordered to the form
\begin{equation}
\frac{1}{2}r^{\lambda }\mathbf{p}^{i}r^{1-2\lambda }\mathbf{p}^{i}r^{\lambda
}=\frac{1}{2}\mathbf{p}^{i}r\mathbf{p}^{i}+\frac{1}{2r}\lambda \left(
\lambda -d+2\right)  \label{ambiguity}
\end{equation}
showing that there is an ordering ambiguity parametrized by $a$ in $%
L_{+^{\prime }0}$. Similarly, there is an ambiguity in $L_{+^{\prime }i}$
parametrized by $b$ as indicated.

The operators $L^{MN}$ should form the algebra of SO$\left( d,2\right) $ in
the quantum theory
\begin{equation}
\left[ L_{MN},L_{RS}\right] =i\eta _{MR}L_{NS}+i\eta _{NS}L_{MR}-i\eta
_{NR}L_{MS}-i\eta _{MS}L_{NR}.
\end{equation}
By using the basic commutation relations among $(\mathbf{r,p)}$ one can
check that the SO$\left( d,2\right) $ commutation relations are indeed
satisfied for any $a,$ and that $b$ is fixed by demanding correct closure
for the commutator
\begin{equation}
\left[ L_{0+^{\prime }},L_{0i}\right] =-iL_{+^{\prime }i},\quad \rightarrow
\quad b=-a-\frac{d-2}{4}.
\end{equation}
The remaining parameter $a$ will be fixed by the Sp$\left( 2\right) $ gauge
invariance, not by the SO$\left( d,2\right) $ algebra, as will be discussed
below.

It is evident that the operators $\mathbf{L}_{ij}$ form the algebra of the
rotation subgroup SO$\left( d-1\right) $. Its quadratic Casimir is given by
\begin{equation}
\mathbf{L}^{2}\equiv \frac{1}{2}L_{ij}L^{ij}=\mathbf{r}^{j}\mathbf{p}^{2}%
\mathbf{r}^{j}-\mathbf{r\cdot p\,p\cdot \,r.}
\end{equation}
Similarly, the following three operators form a SO$\left( 1,2\right) $
subalgebra
\begin{eqnarray}
L_{+^{\prime }-^{\prime }} &\equiv &J_{2},\quad L_{+^{\prime }0}\equiv \frac{%
1}{2}\left( J_{0}+J_{1}\right) ,\quad L_{-^{\prime }0}\equiv J_{0}-J_{1}, \\
J_{2} &=&\frac{1}{2}\left( \mathbf{r\cdot p+p\cdot r}\right) ,\quad \left(
J_{0}+J_{1}\right) =\mathbf{p}^{i}r\mathbf{p}^{i}+\frac{2a}{r},\quad
J_{0}-J_{1}=r
\end{eqnarray}
For any anomaly coefficient $a$ they close correctly
\begin{equation}
\left[ J_{0},J_{1}\right] =iJ_{2},\quad \left[ J_{0},J_{2}\right]
=-iJ_{1},\quad \left[ J_{1},J_{2}\right] =-iJ_{0}.
\end{equation}
The compact generator $J_{0}$ is given in terms of the canonical operators
as
\begin{equation}
J_{0}=L_{+^{\prime }0}+\frac{1}{2}L_{-^{\prime }0}=\frac{1}{2}\mathbf{p}^{i}r%
\mathbf{p}^{i}+\frac{a}{r}+\frac{r}{2}.  \label{compact}
\end{equation}
The quadratic Casimir operator for this subalgebra is
\begin{eqnarray}
j(j+1) &=&J_{0}^{2}-J_{1}^{2}-J_{2}^{2}  \label{related} \\
&=&L_{+^{\prime }0}L_{-^{\prime }0}+L_{-^{\prime }0}L_{+^{\prime }0}-\left(
L_{+^{\prime }-^{\prime }}\right) ^{2} \\
&=&\mathbf{L}^{2}+\frac{1}{4}\left( d-1\right) ^{2}-\frac{1}{2}\left(
d-1\right) +2a \\
&=&\mathbf{L}^{2}+\frac{1}{4}\left( d-2\right) ^{2}-\frac{1}{4}+2a
\end{eqnarray}
We see that the quadratic Casimir operators of the SO$\left( 1,2\right) $
subalgebra and that of the rotation subgroup SO$\left( d-1\right) $ are
related to each other in this representation of SO$\left( d,2\right) $. The
overall quadratic Casimir operator for SO$\left( d,2\right) $ may now be
evaluated
\begin{eqnarray}
C_{2} &=&\frac{1}{2}L_{MN}L^{MN} \\
&=&-\left( L_{+^{\prime }-^{\prime }}\right) ^{2}+L_{+^{\prime
}0}L_{-^{\prime }0}+L_{-^{\prime }0}L_{+^{\prime }0} \\
&&-L_{+^{\prime }i}L_{-^{\prime }i}-L_{-^{\prime }i}L_{+^{\prime
}i}-L_{0i}L_{0i}+\frac{1}{2}L_{ij}L^{ij} \\
&=&\mathbf{L}^{2}+\frac{1}{4}\left( d-2\right) ^{2}-\frac{1}{4}+2a \\
&&-2\mathbf{L}^{2}-\frac{1}{2}\left( d-2\right) ^{2}+2a-\frac{5}{4}+\mathbf{L%
}^{2} \\
&=&-\frac{d^{2}}{4}+d-\frac{3}{2}+4a \\
&=&1-\frac{d^{2}}{4}
\end{eqnarray}
In the last line we required a definite value for the SO$\left( d,2\right) $
Casimir, $C_{2}=1-\frac{d^{2}}{4},$ because this is equivalent to requiring
Sp$\left( 2\right) $ gauge singlets, thus insuring that the states are
physical. The last step fixes the values of $a$ and $b$ uniquely in the
gauge invariant sector
\begin{equation}
a=\frac{1}{8}\left( 5-2d\right) ,\quad b=-\frac{1}{8}.
\end{equation}
These values correspond to an interesting resolution of the quantum ordering
ambiguity (\ref{ambiguity}) of the operators in $L_{+^{\prime }0}$
\[
L_{+^{\prime }0}=\frac{1}{2}\mathbf{p}^{i}r\mathbf{p}^{i}+\frac{1}{8r}\left(
5-2d\right) =r^{\frac{1}{2}}\left[ \frac{1}{2}\mathbf{p}^{2}\right] r^{\frac{%
1}{2}}.
\]

A basis for the quantum theory is chosen to diagonalize the Hamiltonian. As
explained earlier, since we have two timelike dimensions the choice of
``time'' corresponds to a choice of Hamiltonian as a linear combination of
the generators of SO$\left( d,2\right) $. One such choice is dual to another
via Sp$\left( 2\right) $ gauge transformations. We now make the following
choice for ``Hamiltonian''
\begin{eqnarray}
h &=&J_0=L_{+^{\prime }0}+\frac 12L_{-^{\prime }0} \\
&=&r^{\frac 12}\left[ \frac 12\mathbf{p}^2+\frac 12\right] r^{\frac 12}
\nonumber
\end{eqnarray}
Since this is a generator of the SO$\left( 1,2\right) $ algebra it is
diagonalized on the usual SO$\left( 1,2\right) $ basis $|jm>$ where $m$ is
the quantized eigenvalue of the compact generator $J_0$. Evidently the
operator $h$ is positive, therefore $m$ can only be positive. This is
possible only in the positive unitary discrete series representation of SO$%
\left( 1,2\right) $ and the spectrum of $m$ must be
\begin{equation}
m=j+1+n_r,\quad n_r=0,1,2,\cdots .
\end{equation}
where, as we will see shortly, the integer $n_r$ will play the role of the
radial quantum number. Let us now show the relation to the Hydrogen atom
Hamiltonian. Applying $h$ on these states we have
\begin{equation}
r^{\frac 12}\left[ \frac 12\mathbf{p}^2+\frac 12\right] r^{\frac
12}|jm>=m|jm>
\end{equation}
Multiplying it with the operator $r^{-\frac 12}$ from the left, this
equation is rewritten as
\begin{equation}
\left[ \frac 12\mathbf{p}^2+\frac 12-\frac mr\right] \left( r^{\frac
12}|jm>\right) =0.
\end{equation}
We now recognize that the states $|\psi _m>=\left( r^{\frac 12}|jm>\right) $
are eigenstates of the Hydrogen atom Hamiltonian. Actually this is a
rescaled form of the standard Hamiltonian equation written in terms of
dimensionful coordinates and momenta $\mathbf{\tilde{r},\tilde{p}}$%
\begin{equation}
\left[ \frac{\mathbf{\tilde{p}}^2}{2M}-\frac \alpha {\tilde{r}}\right] |\psi
_m>=E_m|\psi _m>.
\end{equation}
The following rescaling relates the two equations and gives the energy of
the atom in terms of the quantum number $m=j+1+n_r$%
\begin{equation}
\mathbf{\tilde{p}=}\frac{M\alpha }m\mathbf{p,\quad }\tilde{r}=\frac
m{M\alpha }r,\quad E_m=-\frac{M\alpha ^2}2\left( j+1+n_r\right) ^{-2}.
\end{equation}

We now give an argument to compute $j$. Since SO$\left( 1,2\right) $
commutes with the SO$\left( d-1\right) $ rotation generators $L_{ij}$, the SO%
$\left( 1,2\right) $ basis can be taken to be simultaneously diagonal with
the SO$\left( d-1\right) $ basis
\begin{equation}
|jml>\sim |SO\left( 1,2\right) ,SO\left( d-1\right) >,
\end{equation}
where $l$ stands for a collection of SO$\left( d-1\right) $ quantum numbers
that we are about to explain. $L_{ij}$ is orbital angular momentum, and its
basis must be constructed by taking direct products of the fundamental unit
vector $\mathbf{\Omega =r/}r$. The only irreducible representations that can
be built in this way are the completely symmetric traceless tensors of SO$%
\left( d-1\right) $. Consider a tensor of rank $l$, i.e. $T_{i_1i_2\cdots
i_l}\left( \mathbf{\Omega }\right) $ which is symmetric and traceless with
indices in $\left( d-1\right) $ dimensions. These provide a complete set of
labels for the states $|jml>$ and are the analogs of the spherical harmonics
in $3$ dimensions. The number of independent components of the tensor in $%
\left( d-1\right) $ space dimensions is
\begin{equation}
N_l\left( d-1\right) =\frac{\left( l+d-4\right) !}{\left( d-3\right) !\,l!}%
\left( 2l+d-3\right)  \label{tensordim}
\end{equation}
This reduces to $\left( 2l+1\right) $ for $d-1=3$, in agreement with
spherical harmonics. The value of the quadratic Casimir of SO$\left(
d-1\right) $ for this representation is
\begin{equation}
\mathbf{L}^2\,|jml>=l\left( l+d-3\right) \,|jml>,\quad l=0,1,2,\cdots
\label{angmom}
\end{equation}
This reduces to $l(l+1)$ for $d-1=3$ in agreement with angular momentum in
three dimensions. Now, we recall that we have established a relation between
the quadratic Casimir operators of SO$\left( 1,2\right) $ and SO$\left(
d-1\right) $ in (\ref{related}). Using this we find
\begin{eqnarray}
j(j+1) &=&l\left( l+d-3\right) +\frac 14\left( d-2\right) \left( d-4\right)
\\
j &=&l+\frac 12\left( d-4\right) .
\end{eqnarray}
Therefore, we have computed the full spectrum. Applying $h$ on these states $%
\left( r^{\frac 12}|jml>\right) $ we now have
\begin{equation}
m=\frac 12\left( d-2\right) +l+n_r
\end{equation}
We may combine the orbital and radial quantum numbers into the total quantum
number as done for the conventional H-atom
\begin{equation}
l+n_r+1=n
\end{equation}
and then write $|jml>=|nl>$ since the complete spectrum depends only on the
total quantum number $n$
\begin{eqnarray}
E_n &=&-\frac{M\alpha ^2}2\left( \frac 12\left( d-4\right) +n\right) ^{-2} \\
n &=&1,2,3,\cdots \\
l &=&0,1,\cdots ,(n-1).
\end{eqnarray}
in agreement with the conventional labelling of the Hydrogen atom states. We
have computed its spectrum in any number of space dimensions $d-1$ and found
that there is a dependence on $d$ in the spectrum: the principal quantum
number $n$ that appears in the denominator is shifted by a half integer $%
\frac 12\left( d-4\right) .$ This shift disappears when $\left( d-1\right)
=3,$ which agrees with the standard result for the Hydrogen atom in three
space dimensions.

We have verified this group theoretical solution by solving directly the Schr%
\"{o}dinger equation for the $1/r$ potential in $D$ space dimensions. The
full wavefunction is $\psi \left( \mathbf{r}\right) =r^{\left( D-1\right)
/2}f\left( r\right) T_{i_1i_2\cdots i_l}\left( \mathbf{\Omega }\right) $,
and the radial equation for any rotationally invariant potential takes the
form
\begin{equation}
\left( -\partial _r^2+\frac 1{r^2}l_D(l_D+1)+v\left( r\right) -\varepsilon
\right) f\left( r\right) =0,
\end{equation}
where $l_D=l+\left( D-3\right) /2$ (try for example the two or three
dimensional cases $D=2,3$). The solution of the radial equation for the $1/r$
$\,$potential proceeds just like the standard three dimensional case, except
for replacing $l_D$ instead of $l$. To compare to our group theoretical
results above we replace $D=d-1.$

We have shown that all the states of the H-atom in $\left( d-1\right) $
dimensions form \textit{a single irreducible representation} of the group SO$%
\left( d,2\right) $ with a quadratic Casimir operator $C_{2}=1-d^{2}/4.$
This group includes a compact group SO$\left( d\right) $ which commutes with
the generator $J_{0}$. Therefore the maximal compact symmetry that commutes
with the Hamiltonian is the rotation group in one more dimension, and the
energy eigenstates remain degenerate under its transformations. This
symmetry is the generalization of the SO$\left( 4\right) $ symmetry of the
H-atom in 3 dimensions. Thus, the two dimensional H-atom has an SO(3)
symmetry, the 3 dimensional H-atom has an SO(4) symmetry, the four
dimensional H-atom has an SO(5) symmetry, and so on. As a consequence of
this symmetry the energy depends only on the total quantum number $n$ and
all the states with different $l=0,1,\cdots ,n-1$ are degenerate. We can
compute this degeneracy $D_{n}$ at a fixed value of $n$ and find
\begin{equation}
D_{n}\left( d-1\right) =\sum_{l=0}^{n-1}N_{l}\left( d-1\right) =\frac{\left(
n+d-3\right) !}{\left( d-2\right) !\,n!}\left( 2n+d-2\right) .
\end{equation}
By comparing to (\ref{tensordim}) we see that it equals the number of
components of a traceless symmetric tensor of rank $n$ in one higher
dimension $D_{n}\left( d-1\right) =N_{n}\left( d\right) $. This is a result
of the SO$\left( d\right) $ symmetry. The computed degeneracy confirms that
indeed the states at a given energy level form a complete multiplet of SO($d$
). The multiplet at a fixed energy level is identified as the completely
symmetric traceless tensor in one more space dimension.

It has been known for a long time that the three dimensional Hydrogen atom
in 3 dimensions has a spectrum that can be described as a representation of
the conformal group SO(4,2) \cite{barut}. We have generalized this result to
any dimension. In comparing the details of our construction to previous work
we find that the details of our construction are somewhat different. Note
especially the issues of ordering of operators at the quantum level. For us
this was crucial from the point of view of Sp$\left( 2\right) $ gauge
invariance and the physical state conditions.

\section{Quantum particle and harmonic oscillator}

Consider the free particle gauge (\ref{relparticle}). The generators of the
conformal group SO$\left( d,2\right) $ are obtained at the classical level
by inserting this gauge into the gauge invariant form (\ref{lmn}). However,
in the quantum theory operator ordering must be taken into account to insure
that all the generators are hermitian and that the algebra of SO$\left(
d,2\right) $ closes correctly. In \cite{dualconf} it was shown that for this
gauge the quantum generators are (at $\tau =0$)
\begin{eqnarray}
L^{ij} &=&\vec{x}^i\vec{p}^j-\vec{x}^j\vec{p}^i \\
L^{+i} &=&-\vec{x}^ip^{+},\quad L^{-i}=x^{-}\vec{p}^i-\frac{\vec{p}^j\vec{x}%
^i\vec{p}^j}{2p^{+}} \\
L^{+-} &=&-\frac 12\left( x^{-}p^{+}+p^{+}x^{-}\right) ,\quad L^{-^{\prime
}+}=\frac 12\vec{x}^2p^{+} \\
L^{+^{\prime }+} &=&p^{+},\quad L^{+^{\prime }-}=\frac{\vec{p}^2}{2p^{+}}%
,\quad L^{+^{\prime }i}=\vec{p}^i \\
L^{+^{\prime }-^{\prime }} &=&\frac 12\left( \vec{x}\cdot \vec{p}\mathbf{+}%
\vec{p}\cdot \vec{x}-x^{-}p^{+}-p^{+}x^{-}\right) \\
L^{-^{\prime }-} &=&\left[
\begin{array}{l}
\frac 1{8p^{+}}\left( \vec{x}^2\vec{p}^2+\vec{p}^2\vec{x}^2-2\alpha \right)
\\
-\frac{x^{-}}2\left( \vec{x}\cdot \vec{p}\mathbf{+}\vec{p}\cdot \vec{x}%
\right) +x^{-}p^{+}x^{-}%
\end{array}
\right] \\
L^{-^{\prime }i} &=&\left[
\begin{array}{l}
\frac 12\vec{x}^j\vec{p}^i\vec{x}^j-\frac 12\vec{x}\cdot \vec{p}\vec{x}^i \\
-\frac 12\vec{x}^i\vec{p}\cdot \vec{x}+\frac 12\mathbf{x}^i\left(
x^{-}p^{+}+p^{+}x^{-}\right)%
\end{array}
\right]
\end{eqnarray}
Hermiticity fixes almost all orders of operators, but the remaining ordering
ambiguity introduces the parameter $\alpha $ in $L^{-^{\prime }-}$ . This is
fixed to $\alpha =-1$ by the commutator $\left[ L^{-^{\prime }i},L^{-\,j}%
\right] =i\delta ^{ij}L^{-^{\prime }-}.$ With this value of $\alpha $ the
quadratic Casimir can then be verified to be precisely $C_2=1-d^2/4$ in
agreement with covariant quantization.

The free particle Hilbert space is defined by diagonalizing the operators $%
p^{+},\vec{p}$, which is the same as diagonalizing the commuting generators $%
L^{+^{\prime }+},L^{+^{\prime }i}.$ The momentum eigenstates $|p^{+},\vec{p}%
> $ form a complete Hilbert space. On this space the free particle
Hamiltonian, which is another generator of the conformal group $L^{+^{\prime
}-}=\frac{\vec{p}^{2}}{2p^{+}},$ is diagonal. These positive norm states
provide a basis for a unitary representation of the conformal group SO$%
\left( d,2\right) $ through the representation of the generators given
above. The Casimir eigenvalues for the representation are fixed as we have
already discussed.

We now show the relation of this representation to the harmonic oscillator.
Instead of diagonalizing $L^{+^{\prime }+},L^{+^{\prime }-},L^{+^{\prime }i}$
we will choose a basis for SO$\left( d,2\right) $ in which the following
operators that commute with each other are simultaneously diagonal
\begin{equation}
L^{ij},L^{+^{\prime }+},\left( L^{+^{\prime }-}+L^{-^{\prime }+}\right) .
\end{equation}
More accurately, only an appropriate commuting subset of orbital angular
momentum operators $L^{ij}$ will be simultaneously diagonal. These operators
correspond to the SO$\left( d-2\right) $ orbital angular momentum $L^{ij}=%
\vec{x}^{i}\vec{p}^{j}-\vec{x}^{j}\vec{p}^{i},$ the lightcone momentum $%
L^{+^{\prime }+}=p^{+}$ and the Hamiltonian of the harmonic oscillator in $%
\left( d-2\right) $ dimensions
\begin{equation}
H\equiv L^{+^{\prime }-}+L^{-^{\prime }+}=\frac{\vec{p}^{2}}{2p^{+}}+\frac{%
p^{+}\vec{x}^{2}}{2},
\end{equation}
where the lightcone momentum $L^{+^{\prime }+}=p^{+}$ in another dimension
plays the role of mass. This choice of Hamiltonian corresponds to another
choice for ``time'', as compared to the choice for ``time'' for the free
particle. The spectrum of this Hamiltonian is well known from the study of
the harmonic oscillator in $d-2$ dimensions
\begin{equation}
E_{n}=n+\frac{1}{2}\left( d-2\right) ,\quad n=0,1,2,\cdots ,
\label{holevels}
\end{equation}
We want to show how this quantum number $n$ and the angular momentum quantum
numbers $l$ $etc.$ are related to the representation space of SO$\left(
d,2\right) .$ To do so, consider the subgroups SO$\left( d-2\right) \otimes $
SO$\left( 2,2\right) $ and label the representation space of SO$\left(
d,2\right) $ by the representations of these subgroups. Recall that $%
SO\left( 2,2\right) =SL\left( 2,R\right) _{L}\otimes SL\left( 2,R\right)
_{R} $. We will show that the Hamiltonian of the harmonic oscillator is the
compact generator of SL$\left( 2,R\right) _{R}$ and that the energy spectrum
of the harmonic oscillator is classified as towers of states corresponding
to the positive discrete series representation $|j_{R},m_{R}>$ of this SL$%
\left( 2,R\right) _{R}$. For every SO$\left( d-2\right) $ angular momentum
quantum number $l$, we will find a relation between $j_{R},j_{L}$ and $l$.

From the general commutation rules for SO$\left( d,2\right) $ one can see
that generators of $SO\left( 2,2\right) =SL\left( 2,R\right) _L\otimes
SL\left( 2,R\right) _R$ are given by
\begin{eqnarray}
G_2^L &=&\frac 12\left( L_{+^{\prime }-^{\prime }}+L_{+-}\right) ,\quad
G_0^L\pm G_1^L=L_{\pm ^{\prime }\pm }\,\,, \\
G_2^R &=&\frac 12\left( L_{+^{\prime }-^{\prime }}-L_{+-}\right) ,\quad
G_0^R\pm G_1^R=L_{\pm ^{\prime }\mp }.
\end{eqnarray}
These satisfy the commutation rules $\left[ G_a^L,G_b^R\right] =0$ and
\begin{eqnarray}
\left[ G_0^{L,R},G_1^{L,R}\right] &=&iG_2^{L,R},\,\,\,\left[
G_0^{L,R},G_2^{L,R}\right] =-iG_1^{L,R}, \\
\left[ G_1^{L,R},G_2^{L,R}\right] &=&-iG_0^{L,R},
\end{eqnarray}
In the present gauge we have the construction
\begin{eqnarray}
G_2^L &=&\frac 14\left( \vec{x}\cdot \vec{p}\mathbf{+}\vec{p}\cdot \vec{x}%
\right) -\frac 12\left( x^{-}p^{+}+p^{+}x^{-}\right) \\
G_0^L+G_1^L &=&p^{+},\quad \\
G_0^L-G_1^L &=&\left[
\begin{array}{l}
\frac 1{8p^{+}}\left( \vec{x}^2\vec{p}^2+\vec{p}^2\vec{x}^2-2\alpha \right)
\\
-\frac{x^{-}}2\left( \vec{x}\cdot \vec{p}\mathbf{+}\vec{p}\cdot \vec{x}%
\right) +x^{-}p^{+}x^{-}%
\end{array}
\right]
\end{eqnarray}
and
\begin{eqnarray}
G_2^R &=&\frac 14\left( \vec{x}\cdot \vec{p}\mathbf{+}\vec{p}\cdot \vec{x}%
\right) ,\quad \\
G_0^R+G_1^R &=&\frac{\vec{p}^2}{2p^{+}},\quad \\
G_0^R-G_1^R &=&\frac 12\vec{x}^2p^{+}.
\end{eqnarray}
Now we see that the Hamiltonian of the harmonic oscillator is the compact
generator of SL$\left( 2,R\right) $%
\begin{equation}
H=2G_{0R}=\frac{\vec{p}^2}{2p^{+}}+\frac{p^{+}\vec{x}^2}2.  \label{hg0}
\end{equation}
In our special representation the quadratic Casimir operators of these
subgroups are related to each other as follows. Defining $%
j_L(j_L+1)=G_{0L}^2-G_{1L}^2-G_{2L}^2$ and $%
j_R(j_R+1)=G_{0R}^2-G_{1R}^2-G_{2R}^2$ we find (for $\alpha =-1$ which
corresponds to $C_2=1-d^2/4$ as seen above)
\begin{equation}
j_R(j_R+1)=j_L(j_L+1)=\frac 14L^2+\frac 1{16}\left( d-2\right) \left(
d-6\right) ,  \label{jljrl}
\end{equation}
where $L^2$ is the quadratic Casimir of SO$\left( d-2\right) $ given by the
quantum ordered form
\begin{equation}
L^2=\vec{p}^i\vec{x}^2\vec{p}^i-\vec{p}\cdot \vec{x}\,\,\vec{x}\cdot \vec{p}.
\label{orbitalcas}
\end{equation}

The unitary representation of SL$\left( 2,R\right) _{R}$ is labelled by $%
|j_{R}m_{R}>$ where $m_{R}$ is the eigenvalue of $G_{0R}$. Since $G_{0R}$ is
a positive operator in our construction, $m_{R}$ can take only positive
values. This is possible only for the positive discrete series
representation, and according to SL$\left( 2,R\right) $ representation
theory it is given by
\begin{equation}
m_{R}=j_{R}+1+n_{r},\quad n_{r}=0,1,2,3,\cdots  \label{mR}
\end{equation}
We will see that the integer $n_{r}\geq 0$ will find an interpretation as
the radial quantum number of the harmonic oscillator. Next we need to find
the allowed values of $j_{R}.$ We saw in eq.(\ref{jljrl}) that $j_{R}$ is
related to angular momentum, therefore we must find the allowed values of
orbital angular momentum SO$\left( d-2\right) $ (\ref{orbitalcas}). As
already explained in the previous section the allowed states for orbital
angular momentum correspond to tensors constructed from the unit vector $%
\vec{\Omega}=\vec{x}/\left| \vec{x}\right| $. The eigenvalues of $L^{2}$ and
the number of states $N_{l}\left( d-2\right) $ are obtained from eqs.(\ref%
{tensordim},\ref{angmom}) by replacing $\left( d-1\right) $ by $\left(
d-2\right) $

\begin{eqnarray}
N_{l}\left( d-2\right) &=&\frac{\left( l+d-5\right) !}{\left( d-4\right)
!\,l!}\left( 2l+d-4\right)  \label{sodm2} \\
L^{2}\,|l &>&=l\left( l+d-4\right) \,|l>,\quad l=0,1,2,\cdots  \nonumber
\end{eqnarray}
Combining eq.(\ref{jljrl}) and eq.(\ref{sodm2}) yields the allowed values of
both $j_{L}$ and $j_{R}$
\[
j_{R}=j_{L}=\frac{1}{2}l+\frac{1}{4}d-\frac{3}{2}.
\]
Inserting this result in (\ref{mR}) one finds
\begin{equation}
m_{R}=\frac{1}{2}l+\frac{1}{4}\left( d-2\right) +n_{r}.
\end{equation}
Now we can compare to the energy spectrum of the harmonic oscillator (\ref%
{holevels}) by using the relation (\ref{hg0}). We see that we must identify $%
l+2n_{r}=n$ where $n$ is the total quantum number and $n_{r}$ is the radial
quantum number in the usual interpretation of the solutions of the Schr\"{o}%
dinger equation. Using the total quantum number $n$ instead of the radial
quantum number $n_{r}$ we summarize our results
\begin{eqnarray}
E_{n} &=&n+\frac{1}{2}(d-2),\,\,\,\quad n=0,1,2,\cdots \\
l &=&n,(n-2),(n-4),\cdots ,(0\,\,or\,\,1) \\
m_{R} &=&\frac{1}{2}n+\frac{1}{4}\left( d-2\right) \\
j_{R} &=&j_{L}=\frac{1}{2}l+\frac{1}{4}d-\frac{3}{2}
\end{eqnarray}
At a fixed energy level $n$ it is well known that the states with different
values of $l$ belong together in SU$\left( d-2\right) $ multiplet
corresponding to the single row Young tableau with $n$ boxes. Instead, here
these states are rearranged vertically as multiplets at the same value of $l$
with different values of the energy $n$. Thus, at each $l$ there is an
SL(2,R)$_{R}$ positive discrete series multiplet $|j_{R},m_{R}>$ which is a
vertical multiplet of different energy levels.

In summary we have found the following labelling of our special
representation by using the harmonic oscillator basis
\begin{eqnarray}
|SO\left( d-2\right) ;SL\left( 2,R\right) _{L};SL\left( 2,R\right) _{R} &>&
\nonumber \\
|l\,\,;j_{L}(l)\,\,p^{+};\,\,j_{R}(l)\,m_{R}(n) &>&
\end{eqnarray}
The $SL\left( 2,R\right) _{L}$ representation is labelled by the eigenvalues
of the generator $G_{0}^{L}+G_{1}^{L}=p^{+}$ which plays the role of mass
for the harmonic oscillator. All the levels taken together make up a single
unitary representation of SO$\left( d,2\right) .$

We have seen that this representation of SO$\left( d,2\right) $ is the same
as the free massless particle representation since it has the same Casimir
eigenvalues. Hence the free massless relativistic particle and the harmonic
oscillator with its mass defined as the lightcone momentum of the particle
are dual to each other in our model.

\section{Summary}

There are two aspects of the model worth emphasizing as potentially more
general than the model itself. One is duality and the other is a larger
covariant space with two timelike dimensions. The concepts of Sp$\left(
2\right) $ duality and two times are inextricably connected to each other in
our model.

As examples of dualities, we have shown that the H-atom, the free particle
and harmonic oscillator are dual to each other. These are some of the
physical systems that can be described by this simple model. A complete
classification of all of its dual sectors has not been obtained at this
stage. At the quantum level the dual sectors are all described by the same
unitary representation of SO$\left( d,2\right) $ with fixed Casimir
eigenvalues. This representation of SO$\left( d,2\right) $ is realized in
terms of different sets of unconstrained canonical variables. In each case a
subset of the SO$\left( d,2\right) $ generators is simultaneously
diagonalized and a particular combination of the generators is interpreted
as the Hamiltonian. Each choice of Hamiltonian corresponds to a fixed gauge
of the duality symmetry in which ``time'' is identified as a particular
combination of the spacetime coordinates which includes two times $%
X^{0^{\prime }},X^{0}$. The topology of the $\left( d+2\right) $ dimensional
spacetime is not the same for each fixed gauge, but each such topology is an
allowed solution of the constraint equations and equations of motion that
follow from a single action $S_{0}$. ``Large'' gauge transformations map the
gauge fixed solutions to each other. This is similar to M-theory dualities
that also map physical systems that live in spaces of different topologies.

Besides dualities, the model shows that familiar physical systems can be
viewed as embedded in a spacetime with two timelike dimensions. This
provides an example for how it is possible to have more than one timelike
dimension and yet describe realistic physics. Furthermore the model shows
that a larger spacetime unifies these physical systems under the same
umbrella.

The duality symmetry in our model is morally similar to the dualities
encountered in M-theory \cite{mtheory}. However, in M-theory the analog of
the action principle that gives rise to dualities remains to be discovered.
It is hoped that our model may provide some new insight into the duality
symmetries in M-theory, and the signals of more than one time or higher
dimensions already noticed from different directions \cite{duff}-\cite%
{sezrudy}.

\end{document}